\documentclass
[nofootinbib,twocolumn,aps,prl,final,nobibnotes]{revtex4}%
\usepackage{amsfonts}
\usepackage{amsmath}
\usepackage{amssymb}
\usepackage{graphicx}
\providecommand{\U}[1]{\protect\rule{.1in}{.1in}}

\begin{document}
\title{Superconductivity in SrTiO$_{3}$: dielectric function method for non-parabolic bands}
\author{S. N. Klimin}
\author{J. Tempere}
\thanks{Also at Lyman Laboratory of Physics, Harvard University, Cambridge, MA 02138, USA.}
\affiliation{Theorie van Kwantumsystemen en Complexe Systemen (TQC), Universiteit
Antwerpen, Universiteitsplein 1, B-2610 Antwerpen, Belgium}
\author{J. T. Devreese}
\affiliation{Theorie van Kwantumsystemen en Complexe Systemen (TQC), Universiteit
Antwerpen, Universiteitsplein 1, B-2610 Antwerpen, Belgium}
\author{J. He}
\author{C. Franchini}
\author{G. Kresse}
\affiliation{Faculty of Physics, Computational Materials Physics, University of Vienna,
Vienna A-1090, Austria}

\begin{abstract}
The dielectric function method for superconductivity has been applied to
SrTiO$_{3}$ accounting for the non-parabolic dispersion of charge carriers in
the conduction band and for the dispersion of optical phonons based on density
functional theory calculations. The obtained critical temperatures of the
superconducting phase transition in SrTiO$_{3}$ are in agreement with
experiments in the density range $n\sim5\times10^{18}$ to $5\times10^{20}$
cm$^{-3}$. The dielectric function method predicts also the sign of the
anomalous isotope effect in strontium titanate, in line with recent observations.

\end{abstract}
\date{\today}
\maketitle

\section{Introduction}

Superconductivity in SrTiO$_{3}$, discovered a long time ago
\cite{Schooley,Koonce,Binnig} attracts a renewed experimental
\cite{Lin,Collignon,Lin2,Rischau,Swartz,Rowley} and theoretical
\cite{Rowley,Ruhman,Rosenstein,JSNM2017,DKTM2014,Edge,Wolfle,Gorkov} interest
because strontium titanate exhibits the superconducting transition at very low
carrier densities.

Different physical mechanisms are considered to explain superconductivity in
SrTiO$_{3}$, e.~g., dynamically screened electron-LO-phonon interactions
\cite{Rowley,Ruhman,Rosenstein,JSNM2017,DKTM2014}, interaction of electrons
with quantum ferroelectric fluctuations \cite{Edge}, with the soft TO-phonon
mode \cite{Wolfle} and with local phonons \cite{Gorkov}. All these studies are
in line with experiments. Nevertheless, the physical picture of
superconductivity in SrTiO$_{3}$ is still far from clarity.

Besides physical mechanisms, the key question concerns the method.
Migdal-Eliashberg theory \cite{McMillan} is hardly applicable here, because
the energies of LO phonons in doped SrTiO$_{3}$ are comparable with the Fermi
energy of electrons at typical concentrations relevant for the experiments.
Nevertheless, this standard approach of superconductivity is exploited in
recent works \cite{Ruhman,Gorkov}. A generalization of the BCS theory beyond
the adiabatic regime was derived by Kirzhnits, Maksimov and Khomskii (KMK)
\cite{Kirzhnits} who described the electron-phonon and Coulomb interactions
through the dielectric function.

As a downside of this generalization, the dielectric function method (DFM)
developed by KMK is a weak-coupling approach. The electron-phonon coupling in
strontium titanate is not weak \cite{GLF}, especially at low concentrations.
Consequently, the DFM can be uncontrolled, especially at low densities (where
the electron-phonon interaction is stronger) and not able to obtain
quantitatively accurate results for SrTiO$_{3}$. However it can give a
qualitatively correct picture of the concentration-dependent superconducting
transition temperature \cite{Rowley,Takada,JSNM2017,DKTM2014}. A verification
of the KMK ansatz within the non-adiabatic extension of the Eliashberg theory
\cite{Rosenstein} showed compatibility of these two methods.

The present work does not claim to explain all aspects of superconductivity in
SrTiO$_{3}$, because the picture seems to be more complicated than can what be
described by a unique mechanism for all concentrations. We restrict the
treatment to only the most usual interactions which are definitely present in
SrTiO$_{3}$: the electron-phonon interaction with LO and acoustic phonons, and
the Coulomb repulsion. The DFM with these interactions seems to be applicable
for SrTiO$_{3}$ except at very low densities $n\lesssim10^{18}%
\operatorname{cm}%
^{-3}$, where the experiment \cite{Lin} reveals a separate superconducting
dome which correlates with the evolution of the Fermi surface with doping
\cite{Collignon2}.

In Refs. \cite{Rowley,Wolfle}, experimental data on the transition temperature
have been reasonably explained using, respectively, DFM and the Morel-Anderson
theory \cite{Morel} in the effective mass approximation, and including in Ref.
\cite{Wolfle} the deformation potential interaction with the soft TO-phonon
mode. In the present work, we apply the extension of the DFM for a
non-parabolic band structure and incorporate the interaction with acoustic
phonons. It is interesting that the softening of the lowest TO-phonon mode
influences the dielectric function and hence enhances $T_{c}$ even before
introducing the deformational electron -- TO-phonon interaction. The other new
element is a multimode dielectric function different from that used in Ref.
\cite{Takada} and accounting for non-parabolic phonon dispersion. We analyze
also the anomalous isotope effect in SrTiO$_{3}$ \cite{Stucky2016}.

\section{Theoretical description}

In this section, DFM is revisited taking into account non-parabolicity of the
conduction band. In the present study, spin-orbit coupling and
next-nearest-neighbor hopping are neglected. We start from the gap equation
\cite{JSNM2017}, where the band dispersion law $\varepsilon_{\mathbf{k}%
,\lambda}$ can be, in general, non-parabolic,%
\begin{align}
\Delta_{\lambda}\left(  \mathbf{k}\right)   &  =-\frac{1}{\left(  2\pi\right)
^{3}}\int d\mathbf{k}^{\prime}~\Delta_{\lambda}\left(  \mathbf{k}^{\prime
}\right)  \frac{\tanh\frac{\beta\left\vert \varepsilon_{\mathbf{k}^{\prime
},\lambda}\right\vert }{2}}{2\left\vert \varepsilon_{\mathbf{k}^{\prime
},\lambda}\right\vert }\nonumber\\
&  \times\frac{2}{\pi}\int_{0}^{\infty}d\Omega\frac{\left\vert \varepsilon
_{\mathbf{k}^{\prime},\lambda}\right\vert +\left\vert \varepsilon
_{\mathbf{k},\lambda}\right\vert }{\Omega^{2}+\left(  \left\vert
\varepsilon_{\mathbf{k}^{\prime},\lambda}\right\vert +\left\vert
\varepsilon_{\mathbf{k},\lambda}\right\vert \right)  ^{2}}\nonumber\\
&  \times V^{R}\left(  \mathbf{k}-\mathbf{k}^{\prime},i\Omega\right)  .
\label{gapeq}%
\end{align}
Here $\beta=1/\left(  k_{B}T\right)  $, $\lambda$ is the index of a subband of
the conduction band, $\Delta_{\lambda}\left(  \mathbf{k}\right)  $ is the
momentum-dependent gap function, $V^{R}\left(  \mathbf{q},i\Omega\right)  $ is
the matrix element of the effective electron-electron interaction in the polar
crystal expressed through the total dielectric function of the electron-phonon
system $\varepsilon^{R}\left(  \mathbf{q},i\Omega\right)  $ in the Matsubara
representation,
\begin{equation}
V^{R}\left(  \mathbf{q},i\Omega\right)  =\frac{4\pi e^{2}}{q^{2}%
\varepsilon^{R}\left(  \mathbf{q},i\Omega\right)  }+V^{ac}\left(
\mathbf{q},i\Omega\right)  , \label{poten}%
\end{equation}
where $V^{ac}\left(  \mathbf{q},i\Omega\right)  $ is the effective potential
due to the acoustic deformation scattering from Ref. \cite{DKTM2014}. The
acoustic-phonon scattering is not a dominating scattering channel in a
strongly polar crystal, but this contribution is taken into account here for completeness.

Next, we use the density-of-states approximation using the density of states
$\nu_{\lambda}\left(  E\right)  $ determined by the equation:%
\begin{equation}
\frac{1}{4\pi^{3}}\int d\mathbf{k}~F_{\lambda}\left(  \varepsilon
_{\mathbf{k},\lambda}\right)  =\int_{\varepsilon_{\lambda,\min}}%
^{\varepsilon_{\lambda,\max}}F_{\lambda}\left(  E\right)  \nu_{\lambda}\left(
E\right)  dE, \label{a1}%
\end{equation}
where $\varepsilon_{\lambda,\max}$ and $\varepsilon_{\lambda,\min}$ are,
respectively, the top and bottom energies in the $\lambda$-th subband, and
$F_{\lambda}\left(  E\right)  $ is an arbitrary function of the energy.

In the density-of-states approximation, the band energy $\varepsilon
_{\mathbf{k},\lambda}$ is modeled by a spherically symmetric band dispersion
$\varepsilon_{k,\lambda}$. The model band dispersion is determined from the
condition that the \emph{density of states for} $\varepsilon_{k,\lambda}$
\emph{is the same as the exact density of states} for a true energy band
dispersion $\varepsilon_{\mathbf{k},\lambda}$. This condition results in the
equation:
\begin{equation}
\int_{\varepsilon_{\lambda,\min}}^{\varepsilon_{\lambda}\left(  k\right)  }%
\nu_{\lambda}\left(  E\right)  dE=\frac{1}{3\pi^{2}}k^{3}. \label{Ek}%
\end{equation}
The root of Eq. (\ref{Ek}) determines the model band energy $\varepsilon
_{\lambda}\left(  k\right)  $ for a given $\nu_{\lambda}\left(  E\right)  $
corresponding to the true band energy $\varepsilon_{\lambda}\left(
\mathbf{k}\right)  $. In this approximation, the gap function depends on the
energy $\varepsilon_{k,\lambda}$, and Eq. (\ref{gapeq}) is transformed to the
equation:%
\begin{align}
\Delta_{\lambda}\left(  \omega\right)   &  =-\int_{\varepsilon_{\lambda,\min
}-\mu}^{\varepsilon_{\lambda,\max}-\mu}\frac{d\omega^{\prime}}{2\omega
^{\prime}}\tanh\left(  \frac{\beta\omega^{\prime}}{2}\right) \nonumber\\
&  \times K_{\lambda}\left(  \omega,\omega^{\prime}\right)  \Delta_{\lambda
}\left(  \omega^{\prime}\right)  , \label{gapeq2}%
\end{align}
with the kernel function for non-parabolic bands:%
\begin{align}
K_{\lambda}\left(  \omega,\omega^{\prime}\right)   &  =\frac{1}{2\pi}\frac
{\nu_{\lambda}\left(  \omega^{\prime}+\mu\right)  }{k_{\lambda}k_{\lambda
}^{\prime}}\int_{\left\vert k_{\lambda}-k_{\lambda}^{\prime}\right\vert
}^{k_{\lambda}+k_{\lambda}^{\prime}}qdq\nonumber\\
&  \times\int_{0}^{\infty}d\Omega\frac{\left\vert \omega^{\prime}\right\vert
+\left\vert \omega\right\vert }{\Omega^{2}+\left(  \left\vert \omega^{\prime
}\right\vert +\left\vert \omega\right\vert \right)  ^{2}}V^{R}\left(
q,i\Omega\right)  , \label{Knp}%
\end{align}
where the energies $\omega,\omega^{\prime}$ are counted from the chemical
potential $\mu$, (which is close to the Fermi energy). The values of the
momentum $k_{\lambda}\equiv p_{\lambda}\left(  \omega\right)  $ and
$k_{\lambda}^{\prime}\equiv p_{\lambda}\left(  \omega^{\prime}\right)  $ are
expressed using the density of states:%
\begin{equation}
p_{\lambda}\left(  \omega\right)  =\left(  3\pi^{2}\int_{\varepsilon
_{\lambda,\min}}^{\mu+\omega}\nu_{\lambda}\left(  E\right)  dE\right)  ^{1/3}.
\label{kE}%
\end{equation}

The total dielectric function entering the effective electron-electron
interaction potential $V^{R}\left(  q,i\Omega\right)  $ is calculated within
the random phase approximation (RPA). In RPA, the total dielectric function is
a sum of the dielectric function of the lattice and the Lindhard polarization
function $P^{\left(  1\right)  }\left(  q,i\Omega\right)  $ for electrons:%
\begin{equation}
\varepsilon^{R}\left(  q,i\Omega\right)  =\varepsilon_{\infty}\prod_{j=1}%
^{n}\left(  \frac{\Omega^{2}+\omega_{L,j}^{2}\left(  q\right)  }{\Omega
^{2}+\omega_{T,j}^{2}\left(  q\right)  }\right)  -\frac{4\pi e^{2}}{q^{2}%
}P^{\left(  1\right)  }\left(  q,i\Omega\right)  . \label{TotEps}%
\end{equation}
The dielectric function of the lattice corresponds to the model of independent
oscillators, where $\omega_{L,j}\left(  q\right)  $ and $\omega_{T,j}\left(
q\right)  $ are, respectively, LO- and TO-phonon frequencies, and
$\varepsilon_{\infty}$ is the high-frequency dielectric constant. Here, we
assume the phonon dispersion law to be isotropic. The TO-mode frequencies have
stronger $q$-dispersion than that for the LO-mode frequencies. Therefore we
suggest only $\omega_{T,j}\left(  q\right)  $ to be $q$-dependent. Here, the
TO-phonon dispersion is modeled by the expression,%
\begin{equation}
\omega_{T,j}\left(  q\right)  =\left[  \omega_{T,j}^{2}+\left(  \omega
_{L,j}^{2}-\omega_{T,j}^{2}\right)  \sin^{2}\left(  qa_{0}/\pi\right)
\right]  ^{1/2}, \label{phondisp}%
\end{equation}
where $a_{0}$ is the lattice constant.

The Lindhard polarization function in the density-of-states approximation
takes the form:%
\begin{align}
P^{\left(  1\right)  }\left(  q,i\Omega\right)   &  =\frac{\pi^{2}}{q}%
\sum_{\lambda}\int_{0}^{\varepsilon_{\lambda,\max}}dE~\frac{\nu_{\lambda
}\left(  E\right)  }{k_{\lambda}\left(  E\right)  }f\left(  E-\mu\right)
\nonumber\\
&  \times\int_{\varepsilon_{\lambda,a}}^{\varepsilon_{\lambda,b}}dE^{\prime
}~\frac{\nu_{\lambda}\left(  E^{\prime}\right)  }{k_{\lambda}\left(
E^{\prime}\right)  }\frac{E-E^{\prime}}{\Omega^{2}+\left(  E-E^{\prime
}\right)  ^{2}}. \label{P}%
\end{align}
with the Fermi distribution functions $f\left(  \varepsilon-\mu\right)  $. The
integration bounds are given by:%
\begin{align}
\varepsilon_{\lambda,a}  &  =\min\left(  \varepsilon_{\lambda}\left(
\left\vert p_{\lambda}\left(  E\right)  -q\right\vert \right)  ,\varepsilon
_{\lambda,\max}\right)  ,\nonumber\\
\varepsilon_{\lambda,b}  &  =\min\left(  \varepsilon_{\lambda}\left(
p_{\lambda}\left(  E\right)  +q\right)  ,\varepsilon_{\lambda,\max}\right)  .
\end{align}

Expanding (\ref{P}) in powers of the momentum up to $q^{2}$, we arrive at the
expansion $\left(  4\pi e^{2}/q^{2}\right)  P^{\left(  1\right)  }\left(
\mathbf{q},i\Omega\right)  =-\omega_{p}^{2}/\Omega^{2}+O\left(  q^{2}\right)
$ with the plasma frequency,%
\begin{equation}
\omega_{p}^{2}=\sum_{\lambda}4\pi e^{2}n_{\lambda}\frac{k_{F,\lambda}}{\pi
^{2}\nu_{\lambda}\left(  \mu\right)  }. \label{plasmf}%
\end{equation}
where $k_{F,\lambda}=\left(  3\pi^{2}n_{\lambda}\right)  ^{1/3}$, and
$n_{\lambda}$ is the population of the $\lambda$-th subband. This
long-wavelength limit allows us to introduce the effective mass parameter
\begin{equation}
m_{\lambda}\equiv\pi^{2}\nu_{\lambda}\left(  \mu\right)  /k_{F,\lambda}
\label{mp}%
\end{equation}
expressed through the density of states at the Fermi energy.

The long-wavelength expansion of $P^{\left(  1\right)  }\left(  \mathbf{q}%
,i\Omega\right)  $ gives a leading contribution to the effective potential.
Therefore the polarization function $P^{\left(  1\right)  }\left(
\mathbf{q},i\Omega\right)  $ calculated using the parabolic band approximation
with the effective mass values given by (\ref{mp}), instead of those for the
bottom of the conduction band, provides a good approximation for the effective
interaction potential $V^{R}\left(  q,i\Omega\right)  $ in non-parabolic
bands. Thus the major effect of the band non-parabolicity comes from the
density of states entering the kernel function (\ref{Knp}) rather than from
the interaction potential.

\section{Concentration dependent critical temperatures}

In this section, we describe the calculated critical temperatures and the
isotope effect in $n$-doped strontium titanate comparing the results with
experimental data and discuss the relation of the present study to other
works. The LO and TO phonon energies are extracted from the results of the
density functional theory calculations using the Vienna ab initio simulation
package (VASP) \cite{Kresse1993,Kresse1996}. All calculations were based on
projector augmented wave pseudopotentials within the density functional
theory. The PBEsol exchange-correlation functional \cite{PBEsol} and a plane
wave energy cutoff of 600 eV were used with a $12\times12\times12$
Monkhorst-Pack $k$-point mesh. The atomic positions were optimized until the
forces were smaller than $0.01$~eV/\AA , and the phonon frequencies were
computed using density functional perturbation theory with an energy accuracy
of $10^{-8}$~eV.

The calculated and measured (shown in brackets) phonon frequencies are shown
in Table \ref{Tab1}. We keep only those LO/TO pairs of frequencies which give
a factor different from in the dielectric function and may be therefore
observed. Even with this reduced set, the DFT calculation gives more
frequencies than measured experimentally, because a significant ratio
$\omega_{L,j}/\omega_{T,j}$ is necessary for detection.%

\begin{table}[h] \centering
\caption{Energies of polar optical phonons in tetragonal SrTiO$_3$ at the Brillouin zone center}%
\begin{tabular}
[c]{|l|c|c|}\hline
No. of the branch & $\hbar\omega_{L,j}$ (meV) & $\hbar\omega_{T,j}$
(meV)\\\hline
1 & 103.62 (98.7 \cite{Gervais93}) & 66.983 (67.6 \cite{VDM2008})\\\hline
2 & 61.521 (58.4 \cite{Gervais93}) & 59.990\\\hline
3 & 54.221 & 52.522\\\hline
4 & 31.637 & 21.40 (21.8 \cite{VDM2008})\\\hline
5 & 18.305 (21.2 \cite{Gervais93}) & 17.767\\\hline
6 & 13.716 & 12.504 (11.0 \cite{VDM2008})\\\hline
\end{tabular}
\label{Tab1}%
\end{table}%

The lowest-energy TO phonon mode, as measured in Ref. \cite{VDM2008}, exhibits
concentration-dependent softening. We use interpolation to these experimental
data for the lowest TO-phonon energy instead of the calculated value from
Table 1 which does not account for softening. The high-frequency dielectric
constant of SrTiO$_{3}$ is chosen according to Refs. \cite{Kamaras,Ruhman},
$\varepsilon_{\infty}=5.1$. The acoustic deformation potential is used
according to Ref. \cite{Janotti}, $D=4$ eV. In order to see the relative
effect of the interaction with acoustic phonons we add also the transition
temperature with $D=0$.%

\begin{figure}[tbh]%
\centering
\includegraphics[
height=2.4033in,
width=3.2067in
]%
{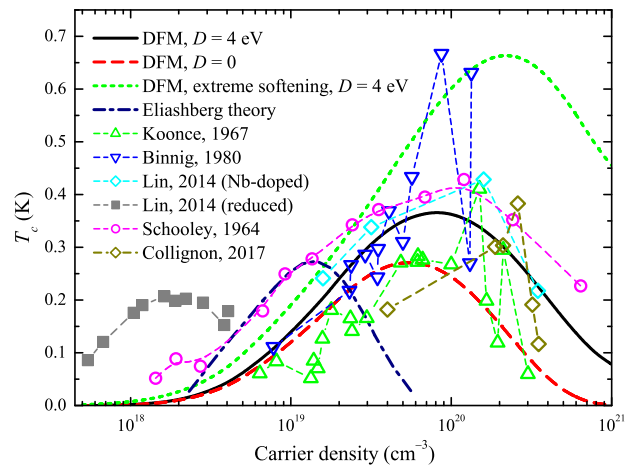}%
\caption{\emph{Solid and dashed curves}: critical temperature in $n$-doped
SrTiO$_{3}$ as a function of the carrier density calculated within DFM using
the density-of-states approximation for non-parabolic bands, with and without
the acoustic-phonon contribution, respectively. \emph{Dotted curve}: the same
assuming the extreme softening of the lowest-energy TO-phonon mode. The
calculated $T_{c}$ is compared with experimental data of Refs.
\cite{Schooley,Koonce,Binnig,Lin,Collignon} (symbols). \emph{Dot-dashed
curve}: the result \cite{Ruhman} of the Eliashberg theory.}%
\end{figure}

The dispersion of the conduction band is described by the tight-binding
analytic fit for the band Hamiltonian using the expressions and notations of
Ref. \cite{JSNM2017}. We apply the values of the diagonal matrix elements
$t_{\delta},t_{\pi}$ from the tight-binding fit to the calculation performed
using the GW method \cite{GW2018}: $t_{\delta}=54.2$ meV, $t_{\pi}=490.9$ meV.
The conduction band splitting is a rather sensitive parameter, because it is
relatively small with respect to diagonal matrix elements. As discussed in
Ref. \cite{Ruhman}, the experimentally relevant values of the interband
splitting are $\delta\varepsilon_{2}\approx2$ meV and $\delta\varepsilon
_{3}\approx8$ meV. These values are obtained when using splitting parameters
$\xi\approx4.615$ meV and $d\approx0.931$ meV.

It is observed in Ref. \cite{Swartz} that the polaronic effect \cite{Varenna}
can substantially influence parameters of the superconducting state and
critical temperatures. It is taken into account, scaling the bare-electron
band energies $\varepsilon_{\mathbf{k},\lambda}$ by $\varepsilon
_{\mathbf{k},\lambda}^{\left(  pol\right)  }=\left(  m_{\lambda}/m_{\lambda
}^{\ast}\right)  \varepsilon_{\mathbf{k},\lambda}$, where the effective mass
parameter $m_{\lambda}$ is determined by (\ref{mp}). In the present
calculation, we have used $\alpha_{eff}\approx2.1$ \cite{KDM2010}, yielding
$m_{\lambda}^{\ast}/m_{\lambda}\approx1.\,35$.

In Fig. 1, we plot critical temperature in $n$-doped SrTiO$_{3}$ as a function
of the carrier density calculated using the dielectric function method within
the density-of-states approximation. The calculated critical temperatures are
compared with experimental data of Refs.
\cite{Schooley,Koonce,Binnig,Lin,Collignon} and with the theoretical result
for $T_{c}$ predicted in Ref. \cite{Ruhman} using the Eliashberg theory. As
can be seen from Fig. 1, the calculated transition temperatures demonstrate
good agreement with the experiments for concentrations $n\sim5\times10^{18}$
to $5\times10^{20}%
\operatorname{cm}%
^{-3}$. A deviation between the present calculation and experiment does not
exceed the deviations between different experiments. The latter is
significant, because very small $T_{c}$ may relatively strongly fluctuate in
different experimental conditions. The present calculation does not match the
experimental data of Refs. \cite{Schooley,Lin} for low concentrations
$n\lesssim10^{18}%
\operatorname{cm}%
^{-3}$, where the calculated transition temperature decreases to small values
with respect to these experimental results. The measured concentration
dependence of $T_{c}$ in this low-density range exhibits a separate dome. The
superconductivity in this low-density regime may be attributed to another
mechanism than the interaction with bulk phonons, for example, local modes
proposed in Ref. \cite{Gorkov}. The incorporation these local modes in the DFM
is beyond the scope of the present work. There are also other pitfalls when
trying to apply DFM at low concentrations: the electron-phonon interaction at
low densities is less screened, so that the weak-coupling approximation may fail.

The non-parabolic dispersion in SrTiO$_{3}$ leads to a rapidly increasing
density of states in the range of energies corresponding to Fermi energies for
densities up to $n\sim10^{20}%
\operatorname{cm}%
^{-3}$. This leads to better quantitative agreement of the calculated $T_{c}$
with experimental data than in our preceding work \cite{JSNM2017}. The
contribution of acoustic phonons results in an increase of $T_{c}$ about 25\%
with a shift of the maximum of superconducting dome to higher densities. The
transition temperature is also calculated assuming extremely strong softening
of the lowest TO-phonon mode to a vanishingly small value. The result is not
sensitive to this soft mode frequency when it is very small. The origin of
this softening can be attributed, e.~g., to quantum criticality \cite{Edge}.
In the present work we do not yet add the deformation potential interaction
with the soft TO-phonon mode considered in Ref. \cite{Wolfle}. Nevertheless,
the softening of TO phonons influences $T_{c}$, because their frequencies
enter the dielectric function and therefore enhance the Fr\"{o}hlich
interaction. As can be seen from Fig. 1, the softening leads to a substantial
rise of $T_{c}$ and to a shift of its maximum to a higher concentration.%

\begin{figure}[tbh]%
\centering
\includegraphics[
height=2.4232in,
width=3.1401in
]%
{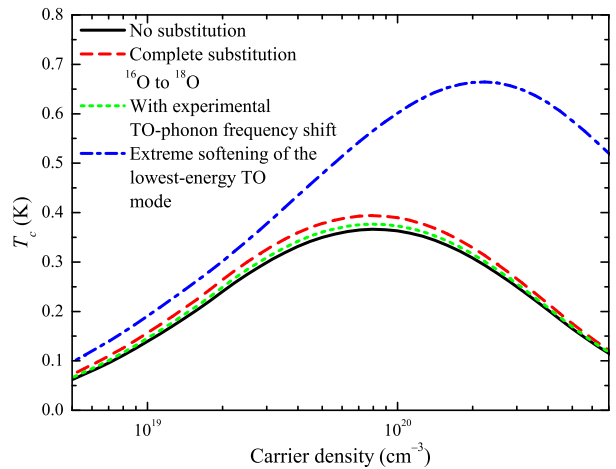}%
\caption{Isotope effect for the critical temperature in $n$-doped SrTiO$_{3}$
calculated using the dielectric function method within the density-of-states
approximation for non-parabolic bands. \emph{Solid curve}: $T_{c}$ without the
isotope substitution. \emph{Dashed curve}: $T_{c}$ with the complete
substitution $^{16}O~\rightarrow~^{18}O$, and assuming proportional
renormalization of all optical phonon frequencies. \emph{Dotted curve}:
$T_{c}$ with the substitution corresponding to the experimental shift of the
TO-phonon frequency. \emph{Dot-dashed curve}: with the extreme softening of
the lowest-energy TO mode.}%
\end{figure}

Fig. 2 shows the calculated isotope effect in SrTiO$_{3}$ calculated using new
band splitting parameters. The curves in Fig. 2 correspond to the following
conditions: (1) without the isotope substitution; (2) with the complete
substitution $^{16}O~\rightarrow~^{18}O$ without assuming appearance of the
soft mode (the continuum approach \cite{Born} gives $\omega_{L\left(
T\right)  ,j}\left(  ^{18}O\right)  /\omega_{L\left(  T\right)  ,j}\left(
^{16}O\right)  \approx0.943$); (3) with the proportional change of phonon
frequencies corresponding to the experimentally observed shift of the highest
TO-phonon energy in Ref. \cite{Stucky2016}; (4) assuming extremely strong
softening of the lowest TO-phonon mode as described above. The softening of
the lowest-energy TO-phonon mode is in line with the observed increase of
$T_{c}$ due to the isotopic substitution $^{16}O~\rightarrow~^{18}O$.

The softening leads to an increase of the effective Fr\"{o}hlich coupling
constant [which in the single-mode picture is proportional to $\left(
1/\varepsilon_{0}-1/\varepsilon_{\infty}\right)  $]. When $\varepsilon
_{0}\rightarrow\infty$, there is an enhancement of $T_{c}$ (see discussions in
\cite{GLF,Rowley}). Also the phonon dispersion favors an increase of the
transition temperature for relatively high concentrations $n\gtrsim10^{19}$
cm$^{-3}$, where the plasmon contribution is not negligible \cite{Takada}. The
plasmon contribution enters $T_{c}$ in a non-additive way, because the
dielectric function within RPA (\ref{TotEps}) leads to a plasmon-phonon
mixing. It is worth noting also that RPA goes beyond the frequently used
plasmon-pole approximation, which results in damping of plasmon-phonon excitations.

\section{Conclusions}

In the present work, we have found a tractable extension of the dielectric
function method for superconductivity to a non-parabolic band structure,
considering the case of SrTiO$_{3}$. The DFM essentially includes retarded
interactions and can be valid in a non-adiabatic regime, restricted by a
weak-coupling approximation.

The density-of-states approach can represent an interest for practical use,
because, first, it allows us to obtain an easily tractable gap equation.
Second, it opens a way to calculate parameters of the superconducting state
when the band dispersion is not precisely given, and only the density of
states is known, e. g., from first-principle calculations or from experimental data.

The dielectric function method is capable to interpret superconductivity in
strontium titanate without adjustment of material parameters, taking them from
experimental data and microscopic calculations. All phonon branches existing
in strontium titanate, as well as the phonon dispersions, are used for the
numerical calculations. Both a non-parabolic band energy and the dispersion of
phonons are essential for quantitative comparison with experiments. The
dielectric function method predicts the sign of the observed unusual isotope
effect in SrTiO$_{3}$ and a rise of the transition temperature due to
softening of the lowest-energy TO-phonon mode. The resulting critical
temperatures are in agreement with experiments in a range of concentrations
$n\sim5\times10^{18}$ to $5\times10^{20}$ cm$^{-3}$. The applicability of DFM
at very low densities remains an open question. An inclusion of other
interactions (e. g., deformation potential interaction with the soft TO-phonon
mode \cite{Wolfle} and the local phonons \cite{Gorkov}) may improve results
for transition temperatures.

The important feature of DFM is a prediction of superconductivity even in the
case when the effective attraction between electrons does not exceed the
Coulomb repulsion in the normal state. As a consequence, the obtained gap
function exhibits a sign reversal near the Fermi surface, being in line with
ab initio microscopic approaches \cite{Sanna}. This makes it interesting to
treat DFM using the electron and phonon densities of states obtained from
first principles.

\begin{acknowledgments}
This work has been supported by the joint FWO-FWF project POLOX (Grant No. I 2460-N36).
\end{acknowledgments}

\end{document}